\documentclass[a4paper,twoside,12pt]{article}

\usepackage{graphicx,amstext,amsmath,amsbsy,amscd}
\usepackage{latexsym}
\usepackage{cite}
\usepackage{graphicx}

\newcommand{\brdot}{^{\scriptscriptstyle{\bullet}}}

\newcommand{\gthree}{{\gamma_\ast}}

\newcommand{\cc}{^*}

\newcommand{\dbsp}{\displaybreak[0]\medsp}

\newcommand{\medsp}{\\[0.7ex]}

\newcommand{\uone}{\mbox{\slshape U}(1)}
\newcommand{\ve}{\varepsilon}

\newcommand{\dega}{\ensuremath{^\dag}}

\newcommand{\diff}[1][]{\mbox{d}#1}

\newcommand{\half}[1]{\ensuremath{\frac{#1}{2}}}
\newcommand{\intd}[1]{\int \!\! #1 \;}
\newcommand{\inv}[1]{\ensuremath{\frac{1}{#1}}}

\newcommand{\Stext}[1]{\itindex{\mathcal{S}}{#1}}

\newcommand{\uhat}{\hat{u}}
\newcommand{\ubhat}{\Hat{\Bar{u}}}

\newcommand{\derfrac}[2][]{\frac{\partial #1}{\partial #2}}

\newcommand{\itindex}[2]{\ensuremath{#1_{\mbox{\scriptsize{\itshape #2}}}}}

\DeclareMathOperator{\hc}{h.c.}

\DeclareMathOperator{\extdm}{d}
\newcommand{\extd}{\extdm \!}

\begin{document}
\renewcommand{\thefootnote}{\fnsymbol{footnote}}
\thispagestyle{empty}
\begin{titlepage}

\begin{flushright}
  TUW-04-05\\
  hep-th/0402138\\
\end{flushright}
\vspace{1cm}

{\centering \textbf{\Large Two-Dimensional $\mathbf{N=(2,2)}$
    Dilaton Supergravity \\ from Graded Poisson-Sigma Models I:}\large
    \par}
\vspace{0.5cm}
{\centering \textbf{\large Complete Actions and Their Symmetries}\par}
\vspace{0.5cm}

\begin{center}
 L. Bergamin\footnote{bergamin@tph.tuwien.ac.at} and W. Kummer\footnote{wkummer@tph.tuwien.ac.at\par}
\end{center}

{\centering \textit{Institute for Theoretical Physics, Vienna University
of Technology}\par}

{\centering \textit{Wiedner Hauptstra{\ss}e 8-10, A-1040 Vienna, Austria}\par}
\vspace{0.5cm}
\begin{abstract}
The formalism of graded Poisson-Sigma models allows the construction
of $N=\left(2,2\right)$ dilaton supergravity in terms of a minimal
number of fields. For the gauged chiral $\uone$ symmetry
the full action, involving all fermionic contributions, is derived.
The twisted chiral case follows by simple redefinition of fields.
The equivalence of our approach to the standard second order one in
terms of superfields is presented, although for the latter so far
only the bosonic part of the action seems to have been available in
the literature. It is shown how ungauged models can be obtained in a
systematic way and some relations to relevant literature in superstring
theory are discussed.
\end{abstract}

\end{titlepage}


\renewcommand{\thefootnote}{\arabic{footnote}}
\setcounter{footnote}{0}
\numberwithin{equation}{section}

\section{Introduction}
\label{sec:introduction}
Motivated mainly by (super-)string theory different versions of $N=\left(2,2\right)$
supergravity in two dimensions \cite{Howe:1987ba,Alnowaiser:1990gh}  have been studied extensively
in superspace
\cite{Gates:1989ey,Gates:1989tn,Ketov:1994tb,Grisaru:1995dr,Grisaru:1995kn,Grisaru:1995dm,Gates:1996du}  some time ago. They also include extensions
to dilaton theory \cite{Nelson:1993vm}.  Actual applications to string theory can
be found more recently in \cite{Gates:2000fj,Haack:2000di,Berkovits:2001tg}. A common drawback of these approaches is the extremely involved formulation,
when the full machinery of (dilaton-) superfields is deployed. For
this reason the results for component expansions obtained so far almost
exclusively are restricted to the bosonic part of those theories.

On the other hand, already in bosonic $D=2$ dilaton gravity the systematic
use of Cartan variables in a {}``temporal'' gauge
\cite{Kummer:1992bg,Kummer:1992rt,Haider:1994cw,Kummer:1995qv} and the subsequent realization that essentially all general dilaton
theories in two dimensions can be interpreted as a special case of
a Poisson-Sigma model (PSM)
\cite{Schaller:1994es,Schaller:1994uj,Ikeda:1993aj,Ikeda:1994fh} have led to
a considerable number of new insights. Not only the extremely simple
derivation of the classical solutions \cite{Kummer:1995qv}, but also
a background independent quantization
\cite{Haider:1994cw,Kummer:1997hy,Kummer:1998zs} has been possible.
For a comprehensive review ref.\ \cite{Grumiller:2002nm}
 can be recommended.

The extension to graded Poisson-Sigma models (gPSMs) has been equally
successful \cite{Ikeda:1994dr,Izquierdo:1998hg,Strobl:1999zz,Ertl:2000si}. As shown by the present authors a certain
subclass of gPSMs, already identified as particularly attractive
from the mathematical point of view in \cite{Bergamin:2002ju} by its dilaton-deformed
super-Poincar\'{e} algebra, could be shown to be equivalent
\cite{Bergamin:2003am} to
the most relevant subclass of 2D dilaton $N=\left(1,1\right)$ supergravity
theories as formulated a long time ago by Park and Strominger
\cite{Park:1993sd}. This permitted the first complete solution (including fermionic fields)
and the formulation of the superpoint particle in a gPSM background
\cite{Bergamin:2003am}, as well as a complete classification of $N=\left(1,1\right)$
solutions retaining certain supersymmetries \cite{Bergamin:2003mh} (BPS states).  In the last
reference also the problem of (non-minimal) coupling of conformal matter to those
supergravities has been solved. Quantization following the same strategy
as in the bosonic case is possible as well \cite{Bergamin:2004us,Bergamin:2004aw}. The much richer structure of extended supergravities, encountered
already in previous work on this subject
\cite{Howe:1987ba,Alnowaiser:1990gh,Gates:1989ey,Gates:1989tn,Ketov:1994tb,Grisaru:1995dr,Grisaru:1995kn,Grisaru:1995dm,Gates:1996du,Nelson:1993vm,Gates:2000fj,Haack:2000di,Berkovits:2001tg}, strongly motivates
the application of gPSM technology to $N=\left(2,2\right).$ As shown
in our present paper this approach indeed is very successful and leads
to novel insights.

In Section \ref{sec:gPSM} we recall the main features of the gPSM formalism, together
with the straightforward implementation of the field content for $N=\left(2,2\right)$
supergravity. Among the two $\uone$ symmetries the gauging
of the chiral or of the twisted chiral case appear as simple alternatives.

Guided by the success of the special {}``dilaton prepotential supergravity''
gPSM in the treatment for $N=\left(1,1\right)$ we immediately concentrate
on that in Section \ref{sec:dpa}---not pursuing the involved elimination process
employed for $N=\left(1,1\right)$ \cite{Bergamin:2002ju}. Indeed already that subclass
of $N=\left(2,2\right)$ gPSMs is eventually found to be equivalent
to the one proposed previously in the superfield approach
\cite{Nelson:1993vm}.

Section \ref{sec:gcs} is devoted to the derivation of the Poisson tensor for
generalized chiral $N=\left(2,2\right)$
gPSM supergravity whereas
in Section \ref{sec:tcs} we show how to reduce the twisted chiral case to the
one of the previous Section by simple redefinition of fields (mirror symmetry).

In Section \ref{sec:actions} we first present the complete $N=\left(2,2\right)$ supergravity
action, involving all fermionic contributions for the chiral case,
with the twisted chiral one to be obtained in an analogous fashion (Section
\ref{sec:actions.1}).
Finally the equivalence to the results obtained in the superfield
formulation \cite{Nelson:1993vm}, where a second order action is used, is the
subject of Section \ref{sec:actions.2}.

In Section \ref{sec:ungauged}
we study the formulation of ungauged supergravity in terms of
gPSMs. Some models are derived explicitely, however interesting questions
about the interpretation of the results cannot be anwered conclusively as yet.

A summary of our results, their relation to previous ones in different
approaches and an outlook concerning some obvious further applications
are contained in Section \ref{sec:conclusions}.
In Appendix \ref{sec:notation} we define our notations.

\section{gPSM for $\mathbf{N=(2,2)}$ supergravity}
\label{sec:gPSM}
A general gPSM consists of scalar fields
$X^I(x)$, which themselves are coordinates of a graded Poisson manifold with
Poisson tensor $P^{IJ}(X) = (-1)^{IJ+1} P^{JI}(X)$. The index
$I$, in the generic case, includes commuting as well as anti-commuting
fields\footnote{The usage of different indices as well as other features of
  our notation are explained in
  Appendix \ref{sec:notation}.
For further details one should consult ref.\ \cite{Ertl:2000si,Ertl:2001sj}.}. In addition one introduces the gauge
potential $A = \extd X^I A_I = \extd X^I A_{mI}(x) \extd x^m$, a one form with respect to the Poisson
structure as well as with respect to the 2d worldsheet coordinates. The gPSM
action reads\footnote{If the multiplication of forms is evident in what
  follows, the wedge symbol will be omitted.}
\begin{equation}
  \label{eq:gPSMaction}
  \begin{split}
    \Stext{gPSM} &= \int_M \extd X^I \wedge A_I + \half{1} P^{IJ} A_J \wedge
    A_I \medsp
    &= \int e \bigl(\partial_0 X^I A_{1 I} - \partial_1 X^I A_{0I} + P^{IJ}
    A_{0J} A_{1I} \bigr) \diff{^2 x}\ .
  \end{split}  
\end{equation}
The Poisson tensor $P^{IJ}$ must have vanishing Nijenhuis tensor (obey a
Jacobi-type identity with respect to the Schouten bracket related as $\{ X^I,
X^J \} = P^{IJ}$ to the Poisson tensor)
\begin{equation}
\label{eq:nijenhuis}
 J^{IJK} =  P^{IL}\partial _{L}P^{JK}+ \mbox{\it 
g-perm}\left( IJK\right) = 0\ ,
\end{equation}
where the sum runs over the graded permutations. The variation of $A_I$ and $X^I$ in \eqref{eq:gPSMaction} yields the gPSM
field equations
\begin{align}
\label{eq:gPSMeom1}
  \extd X^I + P^{IJ} A_J &= 0\ ,\medsp
\label{eq:gPSMeom2}
  \extd A_I + \half{1} (\partial_I P^{JK}) A_K A_J &= 0\ .
\end{align}
Due to
\eqref{eq:nijenhuis} the action \eqref{eq:gPSMaction} is invariant under the
symmetry transformations
\begin{align}
\label{eq:symtrans}
  \delta X^{I} &= P^{IJ} \ve _{J}\ , & \delta A_{I} &= -\mbox{d} \ve
  _{I}-\left( \partial _{I}P^{JK}\right) \ve _{K}\, A_{J}\ ,
\end{align}
where the term $\extd \epsilon_I$ in the second of these equations provides
the justification for calling $A_I$ ``gauge fields''.

For a generic (g)PSM the commutator of two transformations \eqref{eq:symtrans} is a
symmetry modulo the equations of motion (e.o.m.-s) in \eqref{eq:gPSMeom1} only:
\begin{align}
\left[\delta_{\ve_1},\delta_{\ve_2}\right]X^I &= \delta_{\ve_3} X^I
\label{eq:psm1.1} \medsp
\left[\delta_{\ve_1},\delta_{\ve_2}\right]A_I &= \delta_{\ve_3} A_I + \left(\extd X^J+
P^{JK}A_K\right)\partial_J \partial_I P^{RS} \ve_{1\,S} \ve_{2\,R} 
\label{eq:psm1.2} 
\end{align}
Here $\ve_3$ is the new symmetry parameter
\begin{equation}
\label{eq:psm2}
\ve_{3\, I} = \partial_I P^{JK} \ve_{1\,K} \ve_{2\,J} +
P^{JK} \bigl( \ve_{1\, K} \partial_J \ve_{2\, I} - \ve_{2\, K} \partial_J
\ve_{1\, I} \bigr)
\end{equation}
and from this equation it is seen that a generic gPSM obeys a non-linear algebra
with structure functions $\partial_I P^{JK}$.

Only for \( P^{IJ} \) linear
in \( X^{I} \) a closed Lie algebra is obtained, and
\eqref{eq:nijenhuis} reduces to the Jacobi identity for the structure
constants of a Lie group. If the Poisson
tensor has a non-vanishing kernel
there exist (one or more) Casimir functions $C(X)$ obeying
\begin{equation}
\label{eq:casimir}
  \{ X^I, C \} = P^{IJ}\derfrac[C]{X^J} = 0\ ,
\end{equation}
which, when the $X^I$ obey the field equations of motion, are constants of
motion.

In an abstract mathematical sense a gPSM is fully determined by the choice of
its target space, i.e.\ the number of (bosonic and fermionic) target space
variables and the number of (bosonic and fermionic) Casimirs. This statement
is equivalent to the (local) existence of Casimir-Darboux coordinates. The
situation in an application to (super-) gravity is less trivial: Here we need
additional structure (a line-element or a point-particle) and global aspects
with respect to that structure become relevant. Thus we cannot
avoid solving the non-linear identity \eqref{eq:nijenhuis} for a particular
Poisson tensor, which describes (super-)gravity in an explicit manner. A possible way
to implement such a constraint in purely bosonic gravity consists in choosing
the target-space variables
\begin{equation}
  \label{eq:psm3}
  X^i = (X^\phi,X^a) = (\phi,X^a)
\end{equation}
and the gauge fields
\begin{equation}
  \label{eq:psm4}
  A_i = (A_\phi,A_a) = (\omega, e_a)
\end{equation}
as the pairs dilaton/spin-connection and auxiliary-vector/zweibein (for a
discussion of possible generalizations see \cite{Strobl:2003kb}). Local
Lorentz invariance then fixes the component
\begin{equation}
  \label{eq:psm5}
  P^{a \phi} = X^b {\epsilon_b}^a
\end{equation}
of the Poisson tensor and the line element follows from the definition of a
symmetric structure $\eta^{ab}$. This concept straightforwardly generalizes to
$N=(1,1)$ supergravity \cite{Ertl:2000si,Bergamin:2002ju}: one adds a pair of Majorana spinors (dilatino and
gravitino) $X^\alpha = \chi^\alpha$ and $A_\alpha = \psi_\alpha$ to
the target space variables and gauge fields resp.\ and one demands
\begin{equation}
  \label{eq:psm6}
  P^{\alpha \phi} = - \half{1} \chi^\beta \gthree{}_\beta{}^\alpha\ .
\end{equation}
The limit of rigid supersymmetry in the flat space constrains the value of the purely fermionic part
of the Poisson tensor:
\begin{equation}
  \label{eq:psm7}
  P^{\alpha \beta} = - 2i X^c \gamma_c^{\alpha \beta} + \mbox{terms $\propto \gthree$}
\end{equation}
A more detailed implementation of this constraint, especially
taking care of eventual singularities at $Y = X^a X_a/2 = 0$, has been discussed in
\cite{Bergamin:2002ju}.

Prepared with this knowledge about simpler models we can outline the principle
steps that lead to a gPSM formulation of $N=(2,2)$ supergravity:
\begin{description}
\item[Choice of the target space:] First of all we have to determine the number
  of fields of our theory or, equivalently, the number of (local) gPSM
  symmetries. Certainly the bosonic variables must include $\phi$ and $X^a$
  from \eqref{eq:psm3}. As we are dealing with two supersymmetries, we need
  two pairs of Majorana dilatini $(\chi^1_\alpha, \chi^2_\alpha)$ and gravitini
  $(\psi^1_\alpha,\psi^2_\alpha)$ which we combine to complex Dirac
  spinors:
  \begin{align}
    \label{eq:psm8}
    \chi_\alpha &= \inv{\sqrt{2}}(\chi^1_\alpha -i \chi^2_\alpha) &
    \psi_\alpha &= \inv{\sqrt{2}} (\psi^1_\alpha + i \psi^2_\alpha)
  \end{align}
  In addition, the $N=(2,2)$ super-algebra has an internal $\uone_V \times
  \uone_A$ symmetry and none \cite{Gates:2000fj}, one or both of these $\uone$ factors can be
  gauged \cite{Howe:1987ba}. Of course, each gauged $\uone$ leads to an
  additional scalar field, appearing as target-space
  variable of the gPSM. Most intuitive is the choice of one gauged $\uone$:
  Beside the fields $X^a$ and $\omega$, which are eliminated to obtain a
  second order formulation (cf.\ \ref{sec:actions.2}), the target space
  variables then describe the field content of a $N=(2,2)$ matter multiplet,
  while the gauge fields can be viewed as the components of the $N=(2,2)$
  multiplet comprising the zweibein $e_a$. Thus we concentrate on that case, which in addition
  has several advantages: In contrast to the case of two gauged $\uone$
  factors, this is an irreducible representation of
  supersymmetry\footnote{Certainly this is not independent of the observation
  that the gPSM fields fit into $N=(2,2)$ multiplets.}. On the other hand, the gauging of
  one factor reduces the number of invariant terms in the Poisson tensor
  which simplifies the highly non-trivial step of finding a solution to the
  condition \eqref{eq:nijenhuis}. Finally, the dilaton supergravity formulated
  previously \cite{Nelson:1993vm} deals with one gauged $\uone$ as well and thus this
  choice will allow a comparison with that work.

  Therefore, an additional pair of bosonic variables $(\pi,B)$ is added to the
  target-space and the gauge-fields resp., which leads to the final set of
  fields
  \begin{align}
    \label{eq:psm9}
    X^I &= (\phi, \pi, X^a, \chi^\alpha, \bar{\chi}^\alpha)\ , & A_I &= (\omega,
    B, e_a, \psi_\alpha, \bar{\psi}_\alpha)\ .
  \end{align}
\item[Symmetry constraints:] The invariance with respect to local Lorentz and
  $B$-gauge symmetry are fully determined by the underlying
  super-algebra. Local Lorentz invariance fixes the $P^{I \phi}$
  components to
  \begin{align}
    \label{eq:psm10}
    P^{a \phi} &= X^b {\epsilon_b}^a\ , & P^{\pi \phi} &= 0\ , &  P^{\alpha
    \phi} &= - \half{1} \chi^\beta \gthree{}_\beta{}^\alpha\ , &
    P^{\bar{\alpha} \phi} = - \half{1} \bar{\chi}^\beta
    \gthree{}_\beta{}^\alpha\ ,
  \end{align}
  in order to create the appropriate covariant derivatives (cf.\
  \eqref{eq:psm14} and \eqref{eq:psm15} below).

  To determine the $P^{I \pi}$ components, a specific choice of gauging must
  be made. We first concentrate on chiral supergravity (gauged $\uone_V$),
  which implies the choice
  \begin{align}
     \label{eq:psm11}
    P^{a \pi} &= 0 , &  P^{\alpha
    \pi} &= - \half{i} \chi^\beta \gthree{}_\beta{}^\alpha\ , &
    P^{\bar{\alpha} \pi} = \half{i} \bar{\chi}^\beta
    \gthree{}_\beta{}^\alpha\ .
  \end{align}
  In Section \ref{sec:tcs} we will show how one obtains other gaugings from this
  specific result. Also the ungauged theory can be considered as a restricted
  version (Section \ref{sec:ungauged}). Finally we implement local supersymmetry in analogy to \eqref{eq:psm7}
  \begin{equation}
    \label{eq:psm12}
    P^{\alpha \bar{\beta}} = - 2i X^c \gamma_c^{\alpha \beta} + \mbox{terms
    $\propto \gthree$}\ .
  \end{equation}
\item[Rigid supersymmetry:] With the result obtained so far the action of an
  $N=(2,2)$ gPSM may be written as
  \begin{equation}
    \label{eq:psm13}
    \mathcal{S} = \int_M \Bigl( \phi \extd \omega + \pi \extd B + X^a D e_a +
    \chi^\alpha D \psi_\alpha + \bar{\chi}^\alpha D \bar{\psi}_\alpha
    + i X^a (\psi \gamma_a \bar{\psi}) + \half{1} \hat{P}^{AB} A_B A_A
    \Bigr)\ ,
  \end{equation}
  where $\hat{P}^{AB}$ is the $A = (a,\alpha,\bar{\alpha})$ part of the
  Poisson tensor without the specific contribution of
  \eqref{eq:psm12}. Setting $\hat{P}^{AB} = 0$ it is found that the remaining
  components with the covariant derivatives
  \begin{align}
    \label{eq:psm14}
    D e_a &= \extd e_a + \omega \epsilon_a{}^b e_b\ , &&\medsp
\label{eq:psm15}
    D \psi_\alpha &= \extd
    \psi_\alpha - \half{1}(\omega + i B) \gthree{}_\alpha{}^\beta \psi_\beta\ , & D \bar{\psi}_\alpha &= \extd
    \bar{\psi}_\alpha - \half{1}(\omega - i B) \gthree{}_\alpha{}^\beta \bar{\psi}_\beta 
  \end{align}
  obey \eqref{eq:nijenhuis} and thus are a gPSM. As may be checked straightforwardly, the structure functions in
  \eqref{eq:psm2} of this Poisson tensor exactly yield rigid $N=(2,2)$
  supersymmetry on flat space including the $\uone_V$ factor of the internal symmetry
  group. This is an important consistency check of the setup discussed so
  far. The necessary step to be performed in the next Section consists in
  finding a Poisson tensor with non-trivial bosonic potential $P^{ab}$ and
  thus referring to a theory of super\emph{gravity}.
\end{description}

\section{$\mathbf{N=(2,2)}$ dilaton prepotential supergravity}
\label{sec:dpa}
To find supergravity models with non-trivial bosonic potential we could
perform similar steps as in \cite{Ertl:2000si}: Those components not yet fixed
by \eqref{eq:psm10}-\eqref{eq:psm12} can be expanded in terms of Lorentz and
$B$-gauge invariant functions and then the non-linear Jacobi identity \eqref{eq:nijenhuis}
is
solved order by order in the fermions $\chi_\alpha$. But, first, it is
already known from the $N=(1,1)$ case that the solution will not be unique for a
given bosonic potential and, second, the expansion shows that the complexity of
this task is most likely almost unmanageable. Fortunately, a different route
suggests itself from the results of the $N=(1,1)$ theories
\cite{Bergamin:2002ju,Bergamin:2003am}: The identity \eqref{eq:nijenhuis} is
solved for a very special (simple) model only, more complicated theories are
found by means of target space diffeomorphisms \cite{Ertl:2000si} that respect
the constraints \eqref{eq:psm10}-\eqref{eq:psm12}, especially conformal
transformations.

As a simplified theory the $N=(2,2)$ version of the model considered in ref.\
\cite{Izquierdo:1998hg} is chosen, cf.\ also Sections (5.4) and (7.4) of
\cite{Ertl:2000si} as well as refs.\ \cite{Bergamin:2002ju,Bergamin:2003am}. The idea is to
minimize in a first step the contributions to torsion, in other words the dependence on
$X^a$. Thus it is demanded that the Poisson tensor is independent of $X^a$
except for the minimal contributions in \eqref{eq:psm10} and
\eqref{eq:psm12}. Then the expansion of the bosonic potential $P^{ab} =
\epsilon^{ab} V$ reduces to
\begin{equation}
  \label{eq:dpa1}
  V = v + \half{1} \chi^2 v_2 + \half{1} \bar{\chi}^2 \bar{v}_2 + \inv{4}
  \chi^2 \bar{\chi}^2 v_4 \ ,
\end{equation}
where the remaining functions depend on $\phi$ and $\pi$ only. For convenience
we introduce the notation $X = \phi+ i \pi$ and thus $v = v(X, \bar{X})$
etc.\footnote{Of course, the action of chiral supergravity is most conveniently written in
  terms of a complex scalar field $X$ and a complex gauge field $\omega + i
  B$. However, we keep $\omega$ and $B$ separate to simplify the
  generalization to different gaugings.} In the mixed component of the Poisson
tensor
\begin{equation}
  \label{eq:dpa2}
  P^{a \alpha} = F_S^a \chi^\alpha + F_P^a (\chi \gthree)^\alpha + F^{ab}
  (\bar{\chi} \gamma_b)^\alpha
\end{equation}
$F_{S}^a$ and $F_{P}^a$ include contributions proportional to $\chi \gamma^a \bar{\chi}$ and $\chi
\gamma^a \gthree \bar{\chi}$ only, which by means of Fierz identities can be
transformed into contributions that already appear in $F^{ab}$. Therefore,
\eqref{eq:dpa2} reduces to
\begin{gather}
  \label{eq:dpa3.1}
  P^{a \alpha} = F^{ab}
  (\bar{\chi} \gamma_b)^\alpha = (F_{(s)} \eta^{ab} + F_{(a)} \epsilon^{ab}) (\bar{\chi}
  \gamma_b)^\alpha\ , \medsp
  \label{eq:dpa3.2}
\begin{align}
  F_{(s)} &= f_{(s)} + \half{1} \chi^2 \tilde{f}_{(s)}\ ,
  & F_{(a)} &= f_{(a)} + \half{1} \chi^2 \tilde{f}_{(a)}\ .
\end{align}
\end{gather}
According to our conventions $P^{a \bar{\alpha}} = - (P^{a
  \alpha})^*$. Finally, the purely fermionic terms must be
considered. $P^{\alpha \bar{\beta}}$ for chiral gaugings of $\uone$ has
no invariant term proportional to $\gthree$, while $P^{\alpha \beta} = U
  \gthree^{\alpha \beta}$ with
\begin{equation}
  \label{eq:dpa4}
  U = u + \half{1} \chi^2 u_2 + \half{1} \bar{\chi}^2 \tilde{u}_2 + \inv{4}
  \chi^2 \bar{\chi}^2 u_4\ .
\end{equation}
Notice that $\tilde{u}_2$ need not be the complex conjugate of $u_2$.

The implementation of the condition $J^{IJK} = 0$ in eq.\ \eqref{eq:nijenhuis} is
a straightforward, but still tedious calculation. All identities with at least one
$\phi$ or $\pi$ are taken into account by the invariant expansions; identities
with an odd number of bosonic indices contribute to even (zero, two, four)
degrees in the number of fermions, the other ones to odd (one, three) degrees.
We introduce the notation $f(\phi,\pi)' =
\partial f(\phi,\pi)/\partial \phi$ and $\dot{f}(\phi,\pi) =
\partial f(\phi,\pi)/\partial \pi$.
The results of that calculation can be summarized as follows:
\begin{description}
\item[order zero:] $J^{abc} = 0$ is the purely bosonic identity and automatically
  satisfied.\newline $J^{a \alpha \beta} = 0$ relates $P^{a \alpha}$ to $P^{\alpha
    \beta}$:
  \begin{align}
    \label{eq:dpa5}
    f_{(s)} &= \frac{i}{4} u'\ , & f_{(a)} &= 0\ .
  \end{align}
  $J^{a \alpha \bar{\beta}} = 0$ expresses the bosonic potential $v$ in terms
  of $u$, which by
\begin{equation}
\label{eq:dpa6}
v = - \inv{8}(\bar{u} u)'
\end{equation}
  defines it to be a prepotential.
\item[order one:] The $\bar{\chi}$ contribution from $J^{ab\alpha}$ yields
\begin{equation}
\label{eq:dpa7}
\bar{v}_2 = \inv{8}u''\ .
\end{equation}
 The $\chi$ contribution of $J^{ab\alpha}$ as well as $J^{\alpha \beta
    \bar{\gamma}}$ constrain the dependence of the prepotential on $\phi$ and
  $\pi$:
\begin{equation}
\label{eq:dpa8.0}
u = u(\phi+i\pi) = u(X)
\end{equation}
Finally, we obtain from $J^{\alpha \beta
    \gamma} = 0$ that $u_2 = \tilde{u}_2 = 0$.
\item[order two:] $J^{abc} = 0$ is again trivial, while $J^{a \alpha \beta}$
  and $J^{a \alpha \bar{\beta}}$ set the higher order contributions of $P^{a
    \alpha}$ to zero: $\tilde{f}_{(s)} = \tilde{f}_{(a)} = 0$.
\item[orders three and four:] The remaining identities are now almost
  trivial. $J^{abc}$ has one non-vanishing term of order three, which tells us
  that $v_4 = 0$. Similarly one gets from $J^{\alpha \beta \bar{\gamma}}$ that
  $u_4 = 0$. All remaining identities are then automatically satisfied.
\end{description}

Putting the pieces together, the Poisson tensor apart from the
components in \eqref{eq:psm10} and \eqref{eq:psm11} becomes
\begin{gather}
\label{eq:dpa8.1}
  P^{ab} = \epsilon^{ab}\bigl( - \inv{8} (\bar{u} u)' + \inv{16} \bar{u}''
  \chi^2 + \inv{16} u'' \bar{\chi}^2 \bigr)\ , \medsp
  \begin{align} P^{a \alpha} &= \frac{i}{4} u' (\bar{\chi} \gamma^a)^\alpha\ , &
  P^{a \bar{\alpha}} &= \frac{i}{4} \bar{u}' (\chi \gamma^a)^\alpha\ ,
  \end{align}\medsp
  P^{\alpha \bar{\beta}} = - 2 i X^a (\gamma_a)^{\alpha \beta}\ ,\medsp\label{eq:dpa8.4}
  \begin{align} P^{\alpha \beta} &= u \gthree^{\alpha \beta}\ , & P^{\bar{\alpha}
    \bar{\beta}} &= \bar{u} \gthree^{\bar{\alpha}
    \bar{\beta}}\ . \end{align}
\end{gather}
The similarity of this tensor to the related model with $N=(1,1)$
supersymmetry (cf.\ eqs.\ (5.34)-(5.36) in \cite{Ertl:2000si}) is obvious.

As mentioned already in Section \ref{sec:gPSM} the knowledge of eventual
Casimir functions \eqref{eq:casimir} is very important. In case of bosonic gravity or $N=(1,1)$
supergravity, the bosonic part of the Poisson tensor has odd dimension and thus
there exists at least one Casimir function. Here the bosonic part has even
dimension, but the symmetry constraints imply that it can never have full
rank. Thus there exist at least two Casimirs. One of them can be chosen as the
$N=(2,2)$ extension of the one present in any PSM gravity model:
\begin{equation}
  \label{eq:dpa10}
  C = 8 Y - \bar{u} u + \half{1} \chi^2 \bar{u}' + \half{1} \bar{\chi}^2 u'
\end{equation}
The second one is related to the new gauge symmetry. As all bosonic fields are
singlets under the $B$-gauge transformation its body is simply $\pi$. Indeed,
a straightforward calculation shows that
\begin{equation}
  \label{eq:dpa11}
  C_\pi = \pi + \frac{i \bar{u}}{4 C} \chi^2 - \frac{i u}{4 C}
  \bar{\chi}^2 - \inv{C} X^a (\chi \gamma_a \gthree \bar{\chi}) 
\end{equation}
commutes in the sense of \eqref{eq:casimir} with all target space variables. For ground-state configurations ($C = 0$ in
\eqref{eq:dpa10}, cf.\ \cite{Bergamin:2003mh}) with non-vanishing fermion fields the form \eqref{eq:dpa11}
of the second Casimir function needs not be well defined. This problem finds a resolution
within the study of the integrability of the theory \cite{Bergamin:neu}.
In certain cases additional (fermionic or bosonic) Casimir
functions can appear (for $N=(1,1)$ cf.\ \cite{Ertl:2000si}).
\section{General chiral supergravity}
\label{sec:gcs}
For many applications the model \eqref{eq:dpa8.1}-\eqref{eq:dpa8.4} is not yet
general enough, as its bosonic potential $v$ has been restricted to be independent of $Y$. In the case
of $N=(1,1)$ supergravity the present authors have found
\cite{Bergamin:2002ju,Bergamin:2003am} that all ``genuine''
supergravities---i.e.\ gPSM theories which obey symmetry restrictions as
pointed out in Section \ref{sec:gPSM}---are obtained from a model of the type
\eqref{eq:dpa8.1}-\eqref{eq:dpa8.4} by the use of a field-dependent conformal
transformation. This concept requires a generalization in the present case,
as the theory depends on a complex scalar $X$ instead of the real dilaton
$\phi$. Thus, we are looking for a target space diffeomorphism (cf.\
\cite{Ertl:2000si})
\begin{equation}
\label{eq:gcs1}
  X^I\ \  \Longrightarrow \ \ \hat{X}^I = \hat{X}^I(X,\bar{X}) =  \hat{X}^I(\phi,\pi)\ ,
\end{equation}
which yields new gauge potentials and a new Poisson tensor
\begin{align}
\label{eq:gcs2}
  \hat{A}_I &= \derfrac[X^J]{\hat{X}^I} A_J\ , &
  \hat{P}^{IJ} &= (-1)^{K(I+1)}(\partial_K \hat{X}^I)
  P^{KL} (\partial_L \hat{X}^J )\ ,
\end{align}
but leaves the symmetry constraints \eqref{eq:psm10}-\eqref{eq:psm12}
invariant. We do not intend to solve this constraint in full generality, but
use the result from \cite{Bergamin:2002ju}, namely that these target space
diffeomorphisms can be interpreted as conformal transformations.

Consider a generalized conformal transformation that depends on a
generic complex function of $Q(\varphi,\pi)$. 
From the constraints that $X^a$ is a real field the transformation
must be of the form
\begin{align}
  \label{eq:pcs3}
  \hat{X} &= X\ , & \hat{X}^a &= e^{-(Q + \bar{Q})/4} X^a\ , & \hat{\chi}^\alpha &=
  e^{-\bar{Q}/4} \chi^\alpha\ , & \Hat{\Bar{\chi}}^\alpha &=   e^{-Q/4}
  \bar{\chi}^\alpha\ .
\end{align}
This transformation leaves the components \eqref{eq:psm10} and
\eqref{eq:psm11} invariant, while $\hat{P}^{\alpha \bar{\beta}}$
in terms of the fields without hats becomes
\begin{multline}
  \label{eq:ug2}
    \hat{P}^{\alpha \bar{\beta}} = e^{-(Q + \bar{Q})/4} P^{\alpha
    \bar{\beta}}\medsp
  + \inv{16} e^{-(Q + \bar{Q})/4} \Bigl( (\bar{\chi} \chi \gthree^{\alpha
    \beta} - \bar{\chi} \gthree \chi \epsilon^{\alpha \beta})(Q'+\bar{Q}' + i
    \dot{Q} - i \dot{\Bar{Q}}) \medsp
    - \bar{\chi} \gamma^a \chi (\gthree \gamma_a)^{\alpha
    \beta} (Q'-\bar{Q}' + i
    \dot{Q} + i \dot{\Bar{Q}}) \Bigr)\ .
\end{multline}
Clearly the last line has to vanish if local supersymmetry shall still be
implemented by \eqref{eq:psm12}. The
possible solutions are $Q(\phi+ i \pi)$ analytic, $Q(\phi)$ real or $Q(\pi)$
imaginary. The first possibility has an analogue in superspace formulation of chiral
$N=(2,2)$ supergravity \cite{Howe:1987ba}: infinitesimal super-Weyl transformations
preserving the constraints may be written as
\begin{align}
\label{eq:ug4}
  E_a^M \delta E_M^b &= \delta_a^b (\Lambda + \bar{\Lambda})\ , & E_\alpha^M
  \delta E_M^\beta &= \delta_\alpha^\beta \bar{\Lambda}\ , & E_a^M \delta
  E_M^{\bar{\beta}} &= i \gamma_a^{\beta \gamma} D_\gamma \Lambda\ ,
\end{align}
with a \emph{chiral} transformation parameter $\Lambda$. Therefore, in
the superspace formulation the gravitino---the lowest component of
$E_m^\alpha$---transforms with a function that depends on the anti-chiral field
$\bar{\Phi} = \phi-i \pi + \ldots$. The remaining two possibilities have no
obvious analogue in superspace, but real $Q(\phi)$ nonetheless is reminiscent of
the case of non-minimally gauged supergravity (eq.\ (28) in \cite{Howe:1987ba}). 

To follow as close as possible the philosophy of the superspace
formulation $Q(X)$ is chosen as an analytic function in $X$, the
remaining possibilities are certainly interesting but we leave their
investigation for future work. This choice in turn implies by \eqref{eq:gcs2}
with $Z = Q'$
\begin{gather}
\label{eq:pcs4}
\begin{align}
  \hat{\omega} &= \omega + \inv{4} \bigl((Z + \bar{Z}) X^b e_b + \bar{Z} \chi
  \psi + Z \bar{\chi} \bar{\psi}  \bigr) \ , & \hat{B} &= B - \frac{i}{4} (\bar{Z} \chi
  \psi - Z \bar{\chi} \bar{\psi})\ , \end{align}\medsp
\label{eq:pcs5}
\begin{align}
  \hat{e}_a &= e^{(Q + \bar{Q})/4} e_a \ , & \hat{\psi}_\alpha &= e^{\bar{Q}/4}
  \psi_\alpha \ , & \Hat{\Bar{\psi}}_\alpha &= e^{Q/4}
  \bar{\psi}_\alpha\ .
\end{align}
\end{gather}
It is now straightforward to derive the new Poisson tensor in terms of the
variables $\hat{X}^I$. By doing this it is found that the prepotential $u(X)$
transforms as
\begin{align}
  \label{eq:pcs6}
  \hat{u} &= e^{-\bar{Q}/2} u\ , & \dot{\hat{u}} &= i (\hat{u}' +  \bar{Z}
  \hat{u})
\end{align}
and thus $\hat{u}$ no longer represents an analytic function.

To economize writing we drop the hats for the fields of the generalized
Poisson tensor in the following and in addition introduce the new functions
\begin{align}
  \label{eq:pcs7}
  w(X) &= \inv{4} e^{\bar{Q}/2} u\ , & W(X,\bar{X}) &= - 2 w \bar{w}\ .
\end{align}
With \eqref{eq:gcs2} the generalized Poisson tensor becomes
\begin{gather}
  \label{eq:pcs8.1}
  P^{ab} = \epsilon^{ab} \Bigl( e^{-(Q + \bar{Q})/2} W' + \half{1} Y (Z +
  \bar{Z}) + \inv{4} \chi^2 e^{-Q/2} \bar{w}'' + \inv{4}\bar{\chi}^2
  e^{-\bar{Q}/2} w'' \Bigr)\ , \medsp
\label{eq:pcs8.2}
  P^{a \alpha} = i e^{-\bar{Q}/2} w'
  (\bar{\chi}\gamma^a)^\alpha - \frac{\bar{Z}}{4} X^b (\chi \gamma_b \gamma^a
  \gthree)^\alpha\ , \medsp 
\label{eq:pcs8.3}
 P^{a \bar{\alpha}} = i e^{-Q/2} \bar{w}'
  (\chi\gamma^a)^\alpha - \frac{Z}{4} X^b (\bar{\chi} \gamma_b \gamma^a
  \gthree)^\alpha\ , \medsp
\label{eq:pcs8.4}
  P^{\alpha \bar{\beta}} = - 2 i X^a (\gamma_a)^{\alpha \beta}\ , \medsp
\label{eq:pcs8.5}
  \begin{align} P^{\alpha \beta} &= (u + \frac{\bar{Z}}{4} \chi^2)
  \gthree^{\alpha \beta}\ , &  P^{\bar{\alpha} \bar{\beta}} &= (\bar{u} + \frac{Z}{4} \bar{\chi}^2)
  \gthree^{\alpha \beta}\ , \end{align}
\end{gather}
yielding the Casimir functions \eqref{eq:dpa10} and \eqref{eq:dpa11} from \eqref{eq:pcs3}
\begin{gather}
\label{eq:pcs9.1}
  C = 8 \bigl(W +e^{(Q + \bar{Q})/2} ( Y + \inv{4} \chi^2 e^{-Q/2} \bar{w}' +
  \inv{4} \bar{\chi}^2 e^{-\bar{Q}/2} w')\bigr)\ , \medsp
\label{eq:pcs9.2}
  C_\pi = \pi + i e^{\bar{Q}/2} \frac{\bar{w}}{C} \chi^2 - e^{Q/2}\frac{w}{C}
  \bar{\chi}^2 - \frac{e^{(Q+\bar{Q})/2}}{C} X^a (\chi \gamma_a \gthree
  \bar{\chi})\ . 
\end{gather}

\section{Twisted-chiral supergravity}
\label{sec:tcs}
To find other types of gaugings of the internal $\uone_V \times \uone_A$ symmetries the
restrictions on allowed target-space diffeomorphisms is partially relaxed: We
still insist that \eqref{eq:psm10}, \eqref{eq:psm12} and the first equation of
\eqref{eq:psm11} remain unchanged, but we allow changes in the second and
third equation of \eqref{eq:psm11}. Nevertheless, we have to assume that the
$P^{\alpha \pi}$ and $P^{\bar{\alpha} \pi}$ components define sensible
covariant derivatives, i.e.\ they remain functions of $\chi^\alpha$ and
$\bar{\chi}^\alpha$ alone. From the covariant derivatives of
$\psi_\alpha$ in chiral spinor components
\begin{align}
  \label{eq:tcs1.1}
  (D \psi)_+ &= \bigl(\extd - \half{1} (\omega + iB) \bigr) \psi_+\ , & (D
  \psi)_- &= \bigl(\extd + \half{1} (\omega + iB) \bigr) \psi_-\ , \medsp
    \label{eq:tcs1.2}
  (D \bar{\psi})_+ &= \bigl(\extd - \half{1} (\omega - iB) \bigr)
  \bar{\psi}_+\ , & (D
  \bar{\psi})_- &= \bigl(\extd + \half{1} (\omega - iB) \bigr) \bar{\psi}_-\ ,
\end{align}
the remaining transformations are obtained by the simple exchange
\begin{align}
  \label{eq:tcs2.1}
  \chi^+ &\longleftrightarrow \bar{\chi}^+\ , & \psi_+ &\longleftrightarrow \bar{\psi}_+
\end{align}
and
\begin{align}
  \label{eq:tcs2.2}
  \chi^- &\longleftrightarrow \bar{\chi}^-\ , & \psi_- &\longleftrightarrow
  \bar{\psi}_-\ .
\end{align}
The combination of the two is obviously trivial, applying one of them yields
the twisted chiral gauging. Notice that the supersymmetry transformation $X^a
(\psi \wedge \gamma_a \psi)$ is invariant under \eqref{eq:tcs2.1} and/or
\eqref{eq:tcs2.2}.

In principle, all formulae of the previous Sections can be taken over together
with the replacements $\chi^- \leftrightarrow \bar{\chi}^-$ and $\psi_-
\leftrightarrow \bar{\psi}_-$ to describe the twisted-chiral version of
supergravity. However, some expressions become rather lengthy. The
twisted-chiral analogue of the Poisson
tensor \eqref{eq:psm11} and \eqref{eq:pcs8.1}-\eqref{eq:pcs8.5} turns out to be
\begin{gather}
  \label{eq:tcs3.1}
  \begin{align}
    P^{a \pi} &= 0\ , &  P^{\alpha
    \pi} &= - \half{i} \chi^\alpha\ , &
    P^{\bar{\alpha} \pi} = \half{i} \bar{\chi}^\alpha\ ,
  \end{align}\dbsp
    \label{eq:tcs3.2}
\begin{split}
  P^{ab} &= \epsilon^{ab} \Bigl( e^{-(Q + \bar{Q})/2} W' + \half{1} Y (Z +
  \bar{Z}) \medsp &\quad + \inv{4} \chi \bar{\chi}( e^{-Q/2} \bar{w}''+ e^{-\bar{Q}/2} w'')
  + \inv{4} \chi \gthree \bar{\chi}( e^{-Q/2} \bar{w}'' -
  e^{-\bar{Q}/2} w'') \Bigr)\ , \end{split} \dbsp
\label{eq:tcs3.3}
\begin{split}
   P^{a \alpha} &= \half{i} (e^{-\bar{Q}/2} w' + e^{-Q/2} \bar{w}')
  (\chi\gamma^a)^\alpha + \half{i} (e^{-\bar{Q}/2} w' - e^{-Q/2} \bar{w}')
  (\chi\gamma^a \gthree)^\alpha\medsp &\quad  - \inv{8}\bigl((\bar{Z}+Z)
  X^b \epsilon_b{}^a + (\bar{Z} - Z) X^a \bigr) \chi^\alpha  - \inv{8}\bigl((\bar{Z}+Z)
  X^a + (\bar{Z} - Z) X^b \epsilon_b{}^a\bigr) (\chi \gthree)^\alpha\ ,
\end{split}\dbsp
\label{eq:tcs3.4}
\begin{split}
P^{\alpha \bar{\beta}} &= - 2i X^a \gamma_a^{\alpha \beta} + \half{1}\bigl( u +
\frac{\bar{Z}}{4}(\chi \bar{\chi} + \chi \gthree \bar{\chi})\bigr) (\gthree -
\epsilon)^{\alpha \beta}\medsp &\quad + \half{1}\bigl( \bar{u} +
\frac{Z}{4}(\chi \bar{\chi} - \chi \gthree \bar{\chi})\bigr) (\gthree +
\epsilon)^{\alpha \beta}\ ,
\end{split}\dbsp
\label{eq:tcs3.5}
P^{\alpha \beta} = P^{\bar{\alpha} \bar{\beta}} = 0\ .
\end{gather}
In analogy to eqs.\ \eqref{eq:pcs8.2}/\eqref{eq:pcs8.3} we have also $P^{a \bar{\alpha}} =
- (P^{a \alpha})^*$. The symbol $\epsilon$ in eq.\ \eqref{eq:tcs3.4} is the
symplectic tensor used to raise spinor indices, $(\gthree \pm \epsilon)/2$ are
the (anti-)chiral projection operators. Finally, the Casimir functions are
obtained as
\begin{gather}
\label{eq:tcs4.1}
  C = 8 \Bigl( W +  e^{(Q + \bar{Q})/2} \bigl( Y + \inv{4} \chi \bar{\chi}( e^{-Q/2} \bar{w}' + e^{-\bar{Q}/2} w')+
  \inv{4} \chi \gthree \bar{\chi} ( e^{-Q/2} \bar{w}' - e^{-\bar{Q}/2} w'
   )\bigr) \Bigr)\ , \medsp
\label{eq:tcs4.2}
  C_\pi = \pi + i \frac{\chi \bar{\chi}}{C}(e^{\bar{Q}/2} \bar{w} - e^{Q/2}
  w) + i \frac{\chi \gthree \bar{\chi}}{C}(e^{\bar{Q}/2} \bar{w} + e^{Q/2}
  w)  - \frac{e^{(Q+\bar{Q})/2}}{C} X^a (\chi \gamma_a \bar{\chi})\ .
\end{gather}

It is important to notice that the discrete transformations \eqref{eq:tcs2.1}
always are defined \emph{globally}, in contrast to the conformal
transformations considered in Section \ref{sec:gcs}. Thus, the chiral
and twisted-chiral models are physically equivalent. This is the
well-known behavior of the geometrical (topological) sector under
mirror symmetry, while propagating matter degrees of freedom in
general are not invariant.

\section{Actions for gauged dilaton supergravity}
\label{sec:actions}
Having found the explicit Poisson tensors that describe
(twisted-)chiral supergravity, we are now ready to write down the
corresponding supergravity actions and their symmetry transformations. Then
these results are compared to the models known in literature, which are
formulated in an (equivalent) second order derivative formulation.
\subsection{gPSM action and its symmetries}
\label{sec:actions.1}
From eq.\ \eqref{eq:gPSMaction} together with
\eqref{eq:pcs8.1}-\eqref{eq:pcs8.5} the full chiral dilaton supergravity action
becomes
\begin{multline}
  \label{eq:act1}
  \Stext{ch} = \int_M \Bigl( \phi\extd \omega + \pi \extd B + X^a De_a + \chi^\alpha D
  \psi_\alpha + \bar{\chi}^\alpha D \bar{\psi_\alpha} \dbsp
  + \epsilon \bigl( \half{1} Y (Z + \bar{Z}) + e^{-(Q + \bar{Q})/2} W' + \inv{4}
  \chi^2 e^{-Q/2} \bar{w}'' + \inv{4} \bar{\chi}^2 e^{-\bar{Q}/2} w''\bigr) \dbsp
  + \frac{\bar{Z}}{4} X^a (\chi \gamma_a \gamma^b e_b \gthree \psi) +
  \frac{Z}{4} X^a (\bar{\chi} \gamma_a \gamma^b e_b \gthree \bar{\psi})
  - i e^{-\bar{Q}/2} w' (\bar{\chi} \gamma^a e_a \psi) - i e^{-Q/2} \bar{w}'
  (\chi \gamma^a e_a \bar{\psi})\dbsp
  + 2 i X^a \bar{\psi}\gamma_a \psi - \half{1} (u + \frac{\bar{Z}}{4} \chi^2)
  \psi \gthree \psi - \half{1} (\bar{u} + \frac{Z}{4} \bar{\chi}^2)
  \bar{\psi}\gthree\bar{\psi} \Bigr)\ .
\end{multline}
The first two terms proportional $\epsilon$, the
two-dimensional volume form, contain the bosonic potential. This action is invariant under local Lorentz
symmetry and $B$-gauge symmetry, which both are realized linearly according to
eqs.\ \eqref{eq:psm10} and \eqref{eq:psm11} together with \eqref{eq:symtrans}. A special field dependent choice
of $\ve_a$ represents 2D diffeomorphisms, which we need not reproduce
explicitly because of the manifest deffeomorphism invariance of \eqref{eq:act1}. More important in different applications are the supersymmetry
transformations. For the target-space variables the result
\begin{gather}
\label{eq:act2.1}
  \begin{align}
    \delta X &= \bar{\chi} \gthree \bar{\ve}\ , & \delta \bar{X} &= \chi
    \gthree \ve\ ,
  \end{align} \medsp
\label{eq:act2.2}
  \delta X^a = - \inv{4} X^b \bigl( \bar{Z} (\chi \gamma_b \gamma^a \gthree
  \ve) + Z (\bar{\chi}\gamma_b \gamma^a \gthree \bar{\ve}) \bigr) + i \bigl(
  e^{- \bar{Q}/2} w' (\bar{\chi} \gamma^a \ve) +  e^{- Q/2} \bar{w}' (\chi
  \gamma^a \bar{\ve})\ ,\medsp
\label{eq:act2.3}  
  \begin{align}
    \delta \chi^\alpha &= 2 i X^a (\bar{\ve} \gamma_a)^\alpha - (u +
    \frac{\bar{Z}}{4} \chi^2) (\ve \gamma^a)^\alpha\ , & \delta
    \bar{\chi}^\alpha &= 2 i X^a (\ve \gamma_a)^\alpha - (\bar{u} +
    \frac{Z}{4} \bar{\chi}^2) (\bar{\ve} \gamma^a)^\alpha\ ,
  \end{align}
\end{gather}
is obtained, while the expressions for the gauge fields are more complicated:
\begin{gather}
  \label{eq:act3.1}
  \begin{split}
    \delta \omega &= \inv{4} X^b \bigl(\bar{Z}' (\chi \gamma_b \gamma^a
    \gthree \ve) + Z' (\bar{\chi} \gamma_b \gamma^a
    \gthree \bar{\ve}) \bigr) -i \bigl( (e^{-\bar{Q}/2} w')' (\bar{\chi}\gamma^a \ve) + (e^{-Q/2}
    \bar{w}')' (\chi\gamma^a \bar{\ve}) \bigr)\medsp
    &\quad - (u' + \frac{\bar{Z}'}{4} \chi^2) \psi \gthree \ve - (\bar{u}' +
    \frac{Z'}{4} \bar{\chi}^2) \bar{\psi}\gthree\bar{\ve}
  \end{split}\dbsp
  \label{eq:act3.2}
  \begin{split}
    \delta B &= \frac{i}{4} X^b \bigl( Z' (\bar{\chi} \gamma_b \gamma^a
    \gthree \bar{\ve}) - \bar{Z}' (\chi \gamma_b \gamma^a
    \gthree \ve) \bigr) -i \bigl( (e^{-\bar{Q}/2} w')\brdot (\bar{\chi}\gamma^a \ve) + (e^{-Q/2}
    \bar{w}')\brdot (\chi\gamma^a \bar{\ve}) \bigr)\medsp
    &\quad - i (u' + \bar{Z} u - \frac{\bar{Z}'}{4} \chi^2) \psi \gthree \ve -
    i (\bar{u}' + Z \bar{u} -
    \frac{Z'}{4} \bar{\chi}^2) \bar{\psi}\gthree\bar{\ve}
  \end{split}\dbsp
  \label{eq:act3.3}
 \delta e_a = \inv{4} \bigl(\bar{Z} (\chi \gamma_a \gamma^b \gthree \ve) + Z
 (\bar{\chi} \gamma_a \gamma^b \gthree \bar{\ve}) \bigr) e_b + 2i (\psi
 \gamma_a \bar{\ve} + \bar{\psi}
 \gamma_a \ve ) \dbsp
  \label{eq:act3.4}
 \delta \psi_\alpha = - (D \ve)_\alpha + \frac{\bar{Z}}{4} X^a
 (\gamma_a\gamma^b\gthree\ve)_\alpha e_b + i e^{-Q/2} \bar{w}' (\gamma^a
 \bar{\ve})_\alpha e_a - \frac{\bar{Z}}{4} \chi_\alpha (\psi \gthree \ve)\dbsp
\label{eq:act3.5}
 \delta \bar{\psi}_\alpha = - (D \bar{\ve})_\alpha + \frac{Z}{4} X^a
 (\gamma_a\gamma^b\gthree\bar{\ve})_\alpha e_b + i e^{-\bar{Q}/2} w' (\gamma^a
 \ve)_\alpha e_a - \frac{Z}{4} \bar{\chi}_\alpha (\bar{\psi} \gthree \bar{\ve})
\end{gather}
In the last two equations under $B$-gauge transformation the symmetry parameter
$\ve_\alpha$ behaves as $\psi_\alpha$ (cf.\
\eqref{eq:psm15}), $\bar{\ve}_\alpha$ as $\bar{\psi}_\alpha$. Again the
similarity of the action \eqref{eq:act1} together with its supersymmetry
transformations \eqref{eq:act3.1}-\eqref{eq:act3.5} and the result obtained in
$N=(1,1)$ supergravity (cf.\ ref.\ \cite{Bergamin:2003am} eqs.\ (18) and (20)-(25)) is immediate. 

Twisted-chiral supergravity follows from the field reflections
\eqref{eq:tcs2.2}. As the corresponding formulae are quite lengthy, but can be
reconstructed easily, we do not reproduce them here.
\subsection{Relation to second-order formulation}
\label{sec:actions.2}
For $N=(1,1)$ dilaton supergravity a very detailed study of the relation
between the gPSM-based (first-order) formulation and the second order
formulation from superspace has been carried out by the present authors in
ref.\ \cite{Bergamin:2003am}. Here we just want to sketch the basic steps for
$N=(2,2)$ supergravity that basically lead to an equivalent result. We
restrict the explicit calculations to the case
of chiral supergravity, the twisted chiral version again follows by a simple
change of variables.

To make contact with the second order formulation of supergravity as it
follows by integrating out auxiliary fields of superspace, also our auxiliary field $X^a$ and the
part of the spin-connection depending on bosonic torsion must be eliminated. The
necessary steps have been worked out for $N = (1,1)$ supergravities
in detail in \cite{Ertl:2000si}, Section 6.3. As the
procedure is independent of the number of target-space variables as well as of
the details of the Poisson tensor, all formulae can be carried over immediately to the present
application.

Variation of the action \eqref{eq:act1} with respect to $X^a$ yields the
torsion equation and is used to eliminate the independent spin connection
according to
\begin{align}
\label{eq:act10.1}
  \omega_a &= e^m_a \omega_m =
  \tilde{\omega}_a - \tilde{\tau}_a\ , \medsp
\label{eq:act10.2}
  \tilde{\omega}_a &= \epsilon^{mn} \partial_n e_{ma} - 2 i \epsilon^{mn} (\bar{\psi}_n
  \gamma_a \psi_m)\ , \medsp
\label{eq:act10.3}
  \tilde{\tau}_a &= - \half{1} (\partial_a \hat{P}^{AB})
  \epsilon^{mn} e_{Bn} e_{Am}\ .
\end{align}
In the last equation $\hat{P}^{AB}$ are the components
\eqref{eq:pcs8.1}-\eqref{eq:pcs8.3} and \eqref{eq:pcs8.5} of the Poisson tensor
without the minimal
torsion contribution \eqref{eq:pcs8.4}, which has been included in the
definition of the supersymmetry covariant spin connection
$\tilde{\omega}$. After replacing the independent spin-connection by
\eqref{eq:act10.1}-\eqref{eq:act10.3} and a subsequent partial integration,
the action can be varied with respect to $X^a$ again. This finally allows to
eliminate that field by a purely algebraic (and even only linear) equation:
\begin{equation}
  \label{eq:act11}
  X^a = - \epsilon^{an} \bigl(\partial_n \phi + \half{1} (\chi \gthree \psi_n)
  + \half{1} (\bar{\chi} \gthree \bar{\psi}_n) \bigr)
\end{equation}
We introduce the curvature scalar and its partners as
\begin{align}
\label{eq:act12.1}
  \tilde{R} &= 2 \ast \extd \tilde{\omega} \ , & \tilde{B} &=
  \ast \extd B \ ,\medsp
   \tilde{\sigma}_\alpha &= \ast (\tilde{D} \psi)_\alpha \ , &
  \Tilde{\bar{\sigma}}_\alpha &= \ast (\tilde{D} \bar{\psi})_\alpha\ ,
\end{align}
where the covariant derivatives $\tilde{D}$ are defined as in \eqref{eq:psm15}
with $\omega$ replaced by $\tilde{\omega}$. After some algebra the second-order
version of the action \eqref{eq:act1} is found:
\begin{multline}
  \label{eq:act13}
  \Stext{ch} = \intd{\diff{^2 x}} e \biggl( \half{1} \tilde{R} \phi + \tilde{B}
  \pi + \chi^\alpha \tilde{\sigma}_\alpha +\bar{\chi}^\alpha
  \Tilde{\Bar{\sigma}}_\alpha - \half{1} (Z + \bar{Z}) \partial^m \phi
  \partial_m \phi + e^{-(Q + \bar{Q})/2} W'\medsp
   + \inv{4} \bigl(
  \chi^2 e^{-Q/2} \bar{w}'' +  \bar{\chi}^2 e^{-\bar{Q}/2} w''\bigr) + i
  \epsilon^{mn} \bigl(e^{-
  \bar{Q}/2} w' (\bar{\chi} \gamma_n \psi_m)+ e^{-
  Q/2} \bar{w}' (\chi \gamma_n \bar{\psi}_m) \bigr)  \medsp
  - \Bigl( \frac{Z}{8} \bigl( \partial^m \phi (\chi \gthree \psi_m) + 2
  \epsilon^{mn} (\partial_n \phi) \bar{\chi} \bar{\psi}_m - \epsilon^{mn}
  (\bar{\chi} \bar{\psi}_n) (\chi \gthree \psi_m)\bigr) + \hc \Bigr) \medsp
  + \inv{32}(Z - \bar{Z})(\chi^2 \psi^m \psi_m - \bar{\chi}^2 \bar{\psi}^m
  \bar{\psi}_m) + \half{1} \epsilon^{mn} \bigl( u (\psi_n \gthree \psi_m) +
  \bar{u} (\bar{\psi}_n \gthree \bar{\psi}_m) \biggr) 
\end{multline}
From the results of \cite{Bergamin:2003am} it is expected that also here this action is
---up to some field redefinitions--- equivalent to the model of
\cite{Nelson:1993vm}, although in that work only the first line of
\eqref{eq:act13} has been worked out explicitly. This is confirmed by several observations:
\begin{enumerate}
\item The bosonic potential of \eqref{eq:act13} is equivalent to the one of
  \cite{Nelson:1993vm}, and both are equivalent to the bosonic potentials of
  the $N= (1,1)$ case.
\item There exists a kinetic term for $\phi$ but not for $\pi$. As pointed out
  in \cite{Bergamin:2003am} this is a consequence of the first order formalism
  in terms of a gPSM. For the same reason, the gPSM based dilaton supergravity
  does not produce a kinetic term for the dilatino. However, by the field
  redefinition
  \begin{align}
\label{eq:act14}
    \underline{\psi}_m^\alpha &= \psi_m^\alpha - \frac{i}{8} \bar{Z} e^a_m
    \epsilon_{ab} (\bar{\chi} \gamma^b)^\alpha\ , &
    \underline{\bar{\psi}}_m^\alpha &= \bar{\psi}_m^\alpha - \frac{i}{8} Z e^a_m
    \epsilon_{ab} (\chi \gamma^b)^\alpha
  \end{align}
  such a term is generated. This redefinition is necessary, as the conformal
  transformation \eqref{eq:pcs3}-\eqref{eq:pcs5} is not equivalent to a
  super-Weyl transformation in superspace \cite{Howe:1987ba}. There beside the
  multiplication with the conformal factor an additional term is needed to
  preserve the torsion constraints (cf.\ \eqref{eq:ug4}). This generates the kinetic term for the
  fermions but does not affect the scalar fields.
\end{enumerate}

If $Z = 0$ the kinetic term of the dilaton disappers. Such models are related
to $N=(2,2)$ supergravity described in terms of a single supergravity
multiplet $(e^a, \psi^\alpha, B)$
\cite{Howe:1987ba,Alnowaiser:1990gh}. Indeed, on a patch
with $u'' \neq 0$ the dilatino
$\chi$ can be eliminated in that case as well. The resulting action
\begin{multline}
\label{eq:act15}
  \Stext{ch} = \intd{\diff{^2 x}} e \biggl( \half{1} \tilde{R} \phi + \tilde{B}
  \pi - \frac{4}{\bar{u}''} \tilde{\sigma}^2 - \frac{4}{u''}
  \Tilde{\Bar{\sigma}}^2 - \inv{8}(\bar{u} u)' + \half{1}\bigl(u \epsilon^{mn} \psi_n \gthree \psi_m + \bar{u}
  \epsilon^{mn} \bar{\psi}_n \gthree \bar{\psi}_m \bigr)\medsp
  - 2i \bigl( \frac{\bar{u}'}{\bar{u}''} \epsilon^{mn} (\tilde{\sigma}
  \gamma_n \bar{\psi}_k) + \frac{u'}{u''} \epsilon^{mn} (\Tilde{\Bar{\sigma}}
  \gamma_n \psi_k) \bigr) + \inv{4} \bigl( \frac{(u')^2}{u''}
  \psi_m \gamma^m \gamma^n \psi_n + \frac{(\bar{u}')^2}{\bar{u}''}
  \bar{\psi}_m \gamma^m \gamma^n \bar{\psi}_n \bigr)\biggr)
\end{multline}
is written in terms of zweibein, gauge-connection and a complex gravitino. The
fields $\phi$ and $\pi$ are connected with the complex auxiliary field from the
superspace approach (cf.\ \cite{Bergamin:2003am} for the $N = (1,1)$ case).

\section{Ungauged supergravity}
\label{sec:ungauged}
Beside the two versions of minimally gauged $N=(2,2)$ supergravity discussed
so far ungauged versions have been found in the context of superstring
compactifications \cite{Gates:2000fj,Haack:2000di,Berkovits:2001tg}. In this
Section it is shown that models of this type can be obtained in a simple way
from the gPSM formulation of the minimally gauged theories.

To formulate an ungauged model we have to get rid of the target space variable
 $\pi$ in the Poisson tensors of Sections \ref{sec:dpa} and
 \ref{sec:gcs}. To illustrate this procedure
the dilaton prepotential supergravity is taken as an example.

It is straightforward to decouple the additional $\uone$ charge by choosing the corresponding Casimir funcition as a new coordinate replacing the coordinate $\pi$ in \eqref{eq:dpa8.1}-\eqref{eq:dpa8.4}. As $\pi$ appears in the prepotential $u(\phi+i \pi)$ and $\bar{u}(\phi-i\pi)$ the relevant replacement is
\begin{align}
\label{eq:ug1.1}
  \begin{split}
    u(\phi+i\pi) &= \uhat + \inv{4C} \uhat' \bigl(\ubhat \chi^2 - \uhat \bar{\chi}^2 + 4i X^a (\chi \gamma_a \gthree \bar{\chi})\bigr) \medsp
    &\quad + \inv{16 C} \chi^2 \bar{\chi}^2 \bigl(\uhat'' + \inv{C} (\uhat \ubhat' - \uhat' \ubhat)\bigr)\ ,
  \end{split}
\end{align}
Here, $\uhat(\phi+i C_\pi)$ and $\ubhat(\phi-i C_\pi)$ are the prepotentials after replacing $\pi$ by the Casimir function $C_\pi$. For later convenience it is worthwhile to look at the appearance of $C$ inside the new prepotential more in detail. Indeed, one has to insert this prepotential into the expression \eqref{eq:dpa10} to compute the remaining conserved quantity. But then $C$ is seen to appear on the right hand side of that equation as well, which could raise the question of the existence of solutions. But this turns out to be a technical problem and does not lead to inconsistencies: $C$ only appears as inverse power in expressions with fermions. By making the split into soul and body $C = C_B + C_S$ a systematic expansion in $C_S$ can be written down and the inverse powers reduce to expressions in $C_B = 8Y - \uhat \ubhat$. The anti-commuting character of the spinors in $C_S$ gaurantee that this procedure stops at some point. In this specific case the result is found to be especially simple: A straightforward calculation shows that all new contributions of order $\propto \chi^2 \bar{\chi}^2$ from the expansion in $C_S$ cancel! Thus we just can replace all Casimir functions in \eqref{eq:ug1.1} by $C_B$ and obtain a closed, but lenghty expression for the remaining Casimir. 

Now it is straightforward to reformulate the Poisson tensor with the new
variable $C_\pi$, which by definition has vanishing components $P^{C_\pi
  I} = \{ C_\pi, X^I \} \equiv 0$. Consequently its gauge potential only appears in the
kinetic term $C_\pi \extd \tilde{A}_\pi$. Once $C_\pi$ is restricted to a
constant this becomes an irrelevant total derivative. Thus we may
simply redefine $\uhat(\phi)$ as a complex prepotential depending on
the dilaton alone  and
drop all reminiscences of $C_\pi$. The remaining components of the Poisson
tensor still obey the non-linear Jacobi identity
\eqref{eq:nijenhuis}. The ensuing gPSM action describes an ungauged version of dilaton supergravity. It is important
to notice that the basic symmetry principles, local Lorentz invariance encoded
in $P^{\phi I}$ and local supersymmetry transformations in $P^{\alpha
  \bar{\beta}}$ remain invariant under this change of variables.

Though the explicit calculation of the Poisson tensor is straightforward the
lenghty expressions are not very illuminating. Instead the expressions
\eqref{eq:dpa8.1}-\eqref{eq:dpa8.4} may be used, if the $u$ and $\bar{u}$ are seen as the
abbreviations for \eqref{eq:ug1.1} and its hermitian conjugate resp. Notice that
derivatives with respect to the dilaton have to be taken inside $C_B$
as well. 

We add some comments on interesting properties and problems of this
model:
\begin{enumerate}
\item It is important to realize that the ungauged model is not
  equivalent to the (twisted-)chiral theories discussed in the
  previous Sections, although the Poisson tensor of the former
  \emph{locally} can be obtained from the latter. There are two
  sources of inequivalence:
  \begin{enumerate}
  \item The replacement of the target-space coordinate $\pi\rightarrow
  C_\pi$ is not defined globally, as can easily be seen from eq.\
  \eqref{eq:tcs4.2}. In particular, the ungauged model only allows for a
  restricted class of solutions with $C=0$ but non-vanishing fermion
  fields. This could have important physical implications as the
  field configurations with $C=0$ are candidates for BPS states \cite{Bergamin:2003mh}.
  \item Though $P^{C_\pi I}  \equiv 0$ after the replacement $\pi\rightarrow
  C_\pi$, the models with and without $C_\pi$ as target-space variable
  are different: The former still consists of all symplectic leaves
  labelled by the value of the Casimir function $C_\pi$, in the latter
  case one has to choose a \emph{fixed} value of $C_\pi$. Despite the fact that the
  specific value of $C_\pi$ is irrelevant it is obvious that the
  ungauged model now only consists of exactly one symplectic leaf of the
  full theory\footnote{Of course, both theories still consist of a
  foliation with respect to $C$, which has been omitted here for
  simplicity.}. We refer to the discussion of dimensionally reduced
  Chern-Simons gravity \cite{Guralnik:2003we,Grumiller:2003ad,Bergamin:2004me}
  as an example of a PSM with $P^{YI} \equiv 0$ in terms of
  ``physical'' coordinates for a specific field $Y$.
  \end{enumerate}
\item Only the dilaton prepotential supergravity has been discussed
  explicitely so far. Obviously, the procedure of decoupling $\pi$ can
  be performed for the general model of Section \ref{sec:gcs} as
  well. Also, the resulting models still follow from a ``conformal
  transformation'' of the simplified model with $Z=0$: Indeed, the
  model with $Z\neq0$ can be obtained by ``recoupling'' of $C_\pi$, a
  subsequent conformal transformation as discussed in Section
  \ref{sec:gcs} and by decoupling $C_\pi$ again. The Casimir function
  $C_\pi$, being invariant under the transformation \eqref{eq:pcs3},
  does not change its value during this procedure and thus we end up
  in the same symplectic leaf with respect to $C_\pi$ as we started
  from. Therefore, it should be possible to circumvent the detour of
  gauging $\pi$ again. Notice however, that this more direct
  transformation cannot be written as a function $Q(\phi)$, solely
  depending on the dilaton. Rather, the analytic function
  $Q(\phi+i\pi)$ in \eqref{eq:pcs3} must be rewritten as a function
  independent of $\pi$ by substituting $\pi \rightarrow C_\pi$. Then
  $Q$ will depend an \emph{all} target-space coordinates of ungauged
  supergravity. Also, the identifications of the new variables
  according to \eqref{eq:pcs3} will become much more
  complicated. Alternatively, one could define conformal
  transformations with a real function $Q(\phi)$ as explained below
  \eqref{eq:ug2}.
\item As in the chiral case we should address mirror symmetry within these
  models. As a consequence of the absence of additional gaugings one
  finds that any transformation of the form
  \begin{align}
    \label{eq:ug10}
    \hat{\chi}^+ &= \cos \alpha^+ \chi^+ + \sin \alpha^+ \bar{\chi}^+\ ,
    & \hat{\chi}^- &= \cos \alpha^- \chi^- + \sin \alpha^-
    \bar{\chi}^-
  \end{align}
  leaves invariant the covariant derivatives as well as local supersymmetry
  transformation. Except for the case $\alpha^\pm = \pi/4$ all transformations \eqref{eq:ug10} are
  defined globally and have no physical influence. Standard mirror symmetry is
  defined as the discrete subgroup $\alpha = \pi/2$. But it may well be that
  for the special case of ungauged supergravity a generalization of type
  \eqref{eq:ug10} exists in superspace as well.
  \item It remains to check whether this model indeed reproduces the result of
\cite{Gates:2000fj} once the Lagrange multipliers $X^a$ and the
torsion dependent part of $\omega$ are eliminated. As can be seen from
\eqref{eq:ug1.1} the ensuing second-order
Lagrangian is very complicated. At this point we encounter an
additional problem, which is generic for all ungauged supergravity
models obtained in this way. Obviously the torsion equation \eqref{eq:act10.3}
now becomes
non-polynomial in $X^a$ due to \eqref{eq:ug1.1} unless the
prepotential is drastically restricted such that all inverse powers of
$C$ cancel. Therefore the ungauged version of supergravity presented
so far will depend in a non-polynmial way on $\partial_m \phi$ in its
second order formulation, though these terms appear in the fermionic
potential only. It is emphasized that this behavior does not influence
any of the mathematical steps used to eliminate the auxiliary
fields. On the contrary, as the elimination procedure does not depend
on the details of the Poisson tensor
\cite{Ertl:2000si,Bergamin:2003am}, the replacement \eqref{eq:ug1.1}
can be made directly in the second order
formulation, if $X^a$ is regarded as an abbreviation for \eqref{eq:act11}.

It may be important to point out that a rescaling of the dilatino
according to $\chi^\alpha \rightarrow \sqrt{C} \chi^\alpha$ does not
remove the inverse powers of $C$. Indeed, to keep local Lorentz
invariance and local supersymmetry in eqs.\ \eqref{eq:psm10} and
\eqref{eq:psm12} $\psi$, $X^a$ and $e_a$ must be rescaled as well. But
then inverse powers of $C$ re-emerge in expressions involving $u$, $u'$ and
$u''$, esp.\ in the bosonic potential.
\end{enumerate}

To summarize it was found that ungauged versions of supergravity can
be obtained straightforwardly from the (twisted-)chiral versions by
decoupling the scalar field $\pi$, the partner of the $\uone$ gauge
field $B$. Nevertheless, symmetry principles from the gPSM formulation
are not as restrictive in this case as in $N=(1,1)$ supergravity and
consequently a large number of globally different version of ungauged
supergravity were found.
Furthermore,
lacking a suitable $x$-space formulation including spinor terms of the
action of \cite{Gates:2000fj}, we do not have any action from superspace at hand that
such a result could be compared with. 

\section{Outlook and conclusions}
\label{sec:conclusions}
The present paper shows that $N=(2,2)$ dilaton supergravity can be
formulated in terms of a graded Poisson-Sigma model. The strategy of the
construction was motivated by our previous works
\cite{Bergamin:2002ju,Bergamin:2003am} on $N = (1,1)$ supergravity: first we
solved the model for a very special case, general theories were found by means
of conformal transformations, which represent a class of target-space
diffeomorphisms in the gPSM. In this way we obtained the explicit Lagrangians for
minimally gauged chiral dilaton supergravity. The twisted-chiral version is
obtained by mirror symmetry \cite{Gates:2000fj}, which allows an
interpretation as a
target-space diffeomorphism as well. Finally it has been outlined how the gPSM
result, which represents a first-order formulation of supergravity with
non-vanishing bosonic torsion, can be transformed into a second-order
formulation. The latter can be compared with earlier works
\cite{Nelson:1993vm,Howe:1987ba,Alnowaiser:1990gh} on (dilaton) supergravity
in superspace. The equivalence of
the bosonic part of the two dilaton supergravity models, namely the one of
ref.\ \cite{Nelson:1993vm} and our result, is obvious, but in contrast to
\cite{Nelson:1993vm} the gPSM framework allows a compact, but explicit
derivation of \emph{all} spinorial terms as well. In addition the two conserved
quantities (Casimir functions) of the theory have been derived explicitly:
one can be chosen as the supersymmetrized version of the standard Casimir
function of dilaton gravity, coinciding with the ADM mass where such a notion
makes sense. The second one
represents the charge of the additional $\uone$ gauge symmetry. As long as the
topological character of the theory is not destroyed by the coupling of matter
fields, these two quantities essentially describe the complete physical content of the
theory.

The result obtained so far motivates numerous applications and
generalizations. The similarity to the $N=(1,1)$
case \cite{Ertl:2000si,Bergamin:2003am,Bergamin:2003mh,Bergamin:2004us,Bergamin:2004aw} suggests that the
advantages of the gPSM framework again enable an exact treatment of the
classical theory. This allows the determination of the complete classical solution of the
model \cite{Bergamin:neu} including all non-trivial fermion contributions. Propagating degrees of freedom may
be added by coupling matter fields (cf.\ \cite{Bergamin:2003mh}). The superspace formulation is certainly
simpler to derive invariant Lagrangians, which then can be adjusted to obtain
the relevant expression in the gPSM framework. Beside the classical
considerations, it must be possible to quantize pure $N = (2,2)$ dilaton
supergravity too in a
non-perturbatively exact way when
formulated as a gPSM. Matter interactions still can be treated perturbatively
(cf.\ \cite{Bergamin:2004us,Bergamin:2004aw} for the $N=(1,1)$ case). As for
$N=(1,1)$ a classification of all BPS states should be possible.

A yet different aspect is the deformation of $N=(2,2)$ dilaton supergravity to
models exhibiting only $N=(1,1)$ invariance. In the limit where the $N=(2,2)$
invariance is recovered solitonic states may appear. Though it is not
straightforward to realize kink solutions with the dilaton alone
\cite{Guralnik:2003we,Grumiller:2003ad} extensions within supergravity are
possible \cite{Bergamin:2004me}. However, the situation could change with $N=(2,2)$
supergravity as the field content encompasses an additional scalar field.

\subsection*{Acknowledgement}
The authors are grateful to D.\ Grumiller, P.\ van Nieuwenhuizen and
especially to E.\ Scheidegger for
interesting discussions. This work has been supported by the project P-16030-N08 of the
Austrian Science Foundation (FWF).

\appendix
\section{Notations and conventions}
\label{sec:notation}
The conventions are identical to
\cite{Ertl:2000si,Ertl:2001sj}, where additional explanations can be found.

Indices chosen from the Latin alphabet are generic (upper case) or (lower
case) refer to commuting objects, Greek indices are anti-commuting ones. Holonomic coordinates
are labeled by $M$, $N$, $O$ etc., anholonomic ones by $A$, $B$, $C$ etc., whereas
$I$, $J$, $K$ etc.\ are general indices of the gPSM:
  \begin{align}
    X^I &= (X^\phi, X^\pi, X^a, X^\alpha, X^{\bar{\alpha}}) = (\phi, \pi, X^a,
    \chi^\alpha , \bar{\chi}^\alpha)\medsp
    A_I &= (A_\phi, A_\pi, A_a, A_\alpha, A_{\bar{\alpha}}) = (\omega, B, e_a,
    \psi_\alpha, \bar{\psi}_\alpha)
  \end{align}

The summation convention is always $NW \rightarrow SE$, e.g.\ for a
fermion $\chi$: $\chi^2 = \chi^\alpha \chi_\alpha$. Our conventions are
arranged in such a way that almost every bosonic expression is transformed
trivially to the graded case when using this summation convention and
replacing commuting indices by general ones. This is possible together with
exterior derivatives acting \emph{from the right}, only. Thus the graded
Leibniz rule is given by
\begin{equation}
  \label{eq:leibniz}
  \mbox{d}\left( AB\right) =A\mbox{d}B+\left( -1\right) ^{B}(\mbox{d}A) B\ .
\end{equation}

In terms of anholonomic indices the metric and the symplectic $2 \times 2$
tensor are defined as
\begin{align}
  \eta_{ab} &= \left( \begin{array}{cc} 1 & 0 \\ 0 & -1
  \end{array} \right)\ , &
  \epsilon_{ab} &= - \epsilon^{ab} = \left( \begin{array}{cc} 0 & 1 \\ -1 & 0
  \end{array} \right)\ , & \epsilon_{\alpha \beta} &= \epsilon^{\alpha \beta} = \left( \begin{array}{cc} 0 & 1 \\ -1 & 0
  \end{array} \right)\ .
\end{align}
The metric in terms of holonomic indices is obtained by $g_{mn} = e_n^b e_m^a
\eta_{ab}$ and for the determinant the standard expression $e = \det e_m^a =
\sqrt{- \det g_{mn}}$ is used. The volume form reads $\epsilon = \half{1}
\epsilon^{ab} e_b \wedge e_a$; by definition $\ast \epsilon = 1$.

The $\gamma$-matrices are used in a chiral representation:
\begin{align}
\label{eq:gammadef}
  {{\gamma^0}_\alpha}^\beta &= \left( \begin{array}{cc} 0 & 1 \\ 1 & 0
  \end{array} \right) & {{\gamma^1}_\alpha}^\beta &= \left( \begin{array}{cc} 0 & 1 \\ -1 & 0
  \end{array} \right) & {{\gthree}_\alpha}^\beta &= {(\gamma^1
    \gamma^0)_\alpha}^\beta = \left( \begin{array}{cc} 1 & 0 \\ 0 & -1
  \end{array} \right)
\end{align}

Covariant derivatives of anholonomic indices with respect to the geometric
variables $e_a = \extd x^m e_{am}$ and $\psi_\alpha = \extd x^m \psi_{\alpha m}$
include the two-dimensional spin-connection one form $\omega^{ab} = \omega
\epsilon^{ab}$. When acting on lower indices the explicit expressions read
($\half{1} \gthree$ is the generator of Lorentz transformations in spinor space):
\begin{align}
\label{eq:A8}
  (D e)_a &= \extd e_a + \omega {\epsilon_a}^b e_b & (D \psi)_\alpha &= \extd
  \psi_\alpha - \half{1} {{\omega \gthree}_\alpha}^\beta \psi_\beta
\end{align}

Dirac conjugation is defined as $\bar{\chi}^\alpha = \chi\dega
\gamma_0$. Written in components of the chiral representation 
\begin{align}
\label{eq:Achi}
  \chi^\alpha &= ( \chi^+, \chi^-)\ , & \chi_\alpha &= \begin{pmatrix} \chi_+ \\
  \chi_- \end{pmatrix}
\end{align}
the relation between upper and lower indices becomes $\chi^+ = \chi_-$,
$\chi^- = - \chi_+$. Dirac conjugation follows as $\bar{\chi}_- =
\chi_-\cc$, $\bar{\chi}_+ = - \chi_+\cc$, i.e.\ for Majorana spinors $\chi_-$ is real while $\chi_+$ is imaginary.

For two gauge-covariant Dirac spinors $\chi_\alpha$ and $\lambda_\alpha$ the
combinations
\begin{alignat}{3}
\label{eq:A10}
\chi \lambda\ , &\qquad&\qquad \chi \gthree \lambda\ , &\qquad&\qquad \bar{\chi} \gamma^a \lambda
\end{alignat}
and their hermitian conjugates are gauge invariant for chiral gaugings, while
\begin{alignat}{3}
\label{eq:A11}
\bar{\chi} \lambda\ , &\qquad&\qquad \bar{\chi} \gthree \lambda\ , &\qquad&\qquad \bar{\chi} \gamma^a \lambda
\end{alignat}
are invariant for twisted-chiral gaugings. Note that in the latter case the
gravitino $\psi_\alpha$ transforms under gauge transformations as
$\bar{\chi}_\alpha$. Thus in eq.\ \eqref{eq:A11} the bilinear invariants of a gravitino and a
dilatino are obtained by substituting $\lambda \rightarrow \bar{\psi}$.

\providecommand{\href}[2]{#2}\begingroup\raggedright\endgroup

\end{document}